\documentclass[twocolumn]{aastex63}
\usepackage{graphicx}
\usepackage{amssymb}
\usepackage{amsmath}
\usepackage{wasysym}
\usepackage{xfrac}
\usepackage{float}
\usepackage{lineno, blindtext}

\accepted{to ApJL, January 6, 2021}
\shortauthors{Doyle et al.}
\shorttitle{Icy exomoons evidenced by spallogenic nuclides in polluted white dwarfs}

\newcommand{\nBe}{n_{\rm Be}}
\newcommand{\nO}{n_{\rm O}}

\begin{document}

\title{ICY EXOMOONS EVIDENCED BY SPALLOGENIC NUCLIDES IN POLLUTED WHITE DWARFS}

\correspondingauthor{Alexandra E. Doyle}
\email{a.doyle@ucla.edu}

\correspondingauthor{Edward D. Young}
\email{eyoung@epss.ucla.edu}

\author{Alexandra E. Doyle}
\affiliation{Earth, Planetary, and Space Sciences\\
University of California, Los Angeles\\
Los Angeles, CA 90095, USA}

\author{Steven J. Desch}
\affiliation{School of Earth and Space Exploration\\
Arizona State University\\
Tempe, AZ 85287, USA}

\author{Edward D. Young}
\affiliation{Earth, Planetary, and Space Sciences\\
University of California, Los Angeles\\
Los Angeles, CA 90095, USA}

\begin{abstract}

We present evidence that excesses in  Be in polluted white dwarfs (WDs) are the result of accretion of icy exomoons that formed in the radiation belts of giant exoplanets. Here we use excess Be in the white dwarf GALEX J2339-0424 as an example. We constrain the parent body abundances of rock-forming elements in GALEX J2339-0424 and show that the over abundance of beryllium in this WD cannot be accounted for by differences in diffusive fluxes through the WD outer envelope nor by chemical fractionations during typical rock-forming processes. We argue instead that the Be was produced by energetic proton irradiation of ice mixed with rock. We demonstrate that the MeV proton fluence required to form the high Be/O ratio in the accreted parent body is consistent with irradiation of ice in the rings of a giant planet within its radiation belt, followed by accretion of the ices to form a moon that is later accreted by the WD. The icy moons of Saturn serve as useful analogs. Our results provide an estimate of spallogenic nuclide excesses in icy moons formed by rings around giant planets in general, including those in the solar system. While excesses in Be have been detected in two polluted WDs to date, including the WD described here, we predict that excesses in the other spallogenic elements Li and B, although more difficult to detect, should also be observed, and that such detections would also indicate pollution by icy exomoons formed in the ring systems of giant planets.
\end{abstract}
   
\section{Introduction}
\begin{table*}
\caption{\label{table}Abundances by number for GALEX J2339-0424 (GALEX 2667197548689621056) from Klein et al. (2021).}
\begin{center}
\begin{tabular}{lcccc}
\hline \hline
\multicolumn{1}{c}{}& \multicolumn{4}{c}{GALEX J2339-0424}\\
$z$ & $n$($z$)/$n$(He) & $\sigma_{\rm spread}$ & $n$($z$)/$n$(Fe) & $\sigma_{\rm spread}$ \\
 &($10^{-8}$)&($10^{-8}$)&&\\
\hline
Be  & 0.0041  & 0.0010   & $3.98 \times 10^{-4}$ & $1.40 \times 10^{-4}$ \\
O   & 298.9   & 25.7     & 29.02                 & 7.74 \\
Mg  & 26.42   & 4.0      & 2.56                  & 0.75 \\
Si  & 25.6    & 4.3      & 2.49                  & 0.75 \\
Ca  & 0.94    & 0.28     & $9.13 \times 10^{-2}$ & $3.56 \times 10^{-2}$ \\
Ti  & 0.027   & 0.007    & $2.62 \times 10^{-3}$ & $9.49 \times 10^{-4}$ \\
V   & $<$0.0046  &   & $< 4.47 \times 10^{-4}$ & \\
Cr  & 0.19    & 0.03     & $1.85 \times 10^{-2}$ & $5.49 \times 10^{-3}$ \\
Mn  & 0.094   & 0.006    & $9.13 \times 10^{-3}$ & $2.38 \times 10^{-3}$ \\
Fe  & 10.3    & 2.6      & 1.00                  & 0.36\\
\end{tabular}
\end{center}
\end{table*}

White dwarfs represent the last stage of stellar evolution. These stellar remnants are extremely dense and have extraordinary gravity such that elements heavier than helium sink rapidly below their surfaces. One would expect to observe only H and He at the surfaces of WDs. However, 25 - 50\% of WDs exhibit elements heavier than helium \citep{RN16, RN17, RN19}. These white dwarfs are `polluted' by heavy elements resulting from accretion of asteroid-like or comet-like bodies \citep{RN7}. Many WDs have observable debris disks from shredded rocky remnants, and a few WDs possess evidence of transiting rocky bodies \citep{Vanderburg2015,Manser_2019,Vanderbosch2019}. The geochemical compositions of extrasolar rocky bodies accreting onto WDs is a burgeoning field unto itself \citep[e.g.,][]{RN10,RN17,RN8,RN1,Vennes2010,Melis2011,Farihi2011,RN481,RN9,RN91,RN51,Hollands_2018,Doyle_2019,Swan_2019,Bonsor2020}.

Recently, exceptionally high and robust over-abundances of Be relative to other rock-forming elements (e.g., Fe, Mg, and O) were discovered in two polluted WDs, GALEX J2339-0424 and GD 378 (Klein et al., 2021), and by inference in the planetary materials polluting them. Lithium (Li), boron (B) and Be share the characteristic of being the products of spallation reactions; all three elements are under-abundant in terms of cosmic abundances but enriched by cosmic rays (CRs) as the result of spallation reactions involving collisions between protons and carbon and oxygen atoms in the interstellar medium (ISM). Rather than being produced by stellar nucleosynthesis, Li is destroyed, or astrated, in stellar interiors at temperatures $> 2.5 \times 10^6$ K, and Be and B are destroyed at temperatures $> 3.5 \times 10^6$ K and $> 5.3 \times 10^6$ K \citep{VangioniFlam2000}. Lithium is easily ionized and thus is more difficult to observe in the optical regime for stars with higher $T_{\rm eff}$, such as GALEX J2339-0424 and GD 378. However, the first detection of Li in polluted WDs was just recently reported for two ultra-cool white dwarfs with $T_{\rm eff} < 4500$K  \citep{Kaiser2020}. The strongest line of boron can be found in the UV (1362.461 $\mathring{\rm A}$) but sufficient data in the UV have not yet been acquired for the polluted WDs discussed here. Nonetheless, the recent detections of Be and Li suggest that a detection of B is highly likely, given sufficient data quality.

Of the isotopes of these elements, only $^7$Li is produced in significant amounts by stellar nucleosynthesis and as a Big-Bang relic \citep[e.g.,][]{Claytonhandbook2003}. The stable isotope of Be is $^{9}$Be and it is produced by the reaction $^{16}$O(p,X)$^{9}$Be where X, the ejected particles, in this case refers to 3p$\alpha$n. We emphasize again that no $^9$Be is produced in stars. Beryllium is a rare element in the Earth’s crust, as well as in the universe, but concentrations of 
the rare radionuclide $^{10}$Be ($t_{1/2} = 1.4$ Myr) in rocks vary with cosmic ray intensities, depth below the surface, and age \citep[e.g.,][]{Somayajulu1977}, affording an age dating technique. On Earth, the formation of primordial and cosmogenic nuclides by cosmic ray spallation occurs in the upper atmosphere, and the radio-isotope products, including $^{10}$Be, that precipitate onto Earth's surface are often used to date deep-sea sediments \citep[e.g][]{Arnold1956,Lal1967} and the recycling of sediments through volcanoes \citep{Morris1990}. An analogous means of using radio-isotope spallation products (collectively referred to as cosmogenic nuclides) has been contemplated for dating the surfaces of icy moons in the solar system \citep{Nordheim2019,Hedman2019}.

In this paper we consider the possible mechanisms for enriching a rocky or icy body in spallogenic nuclides. We consider various sources of MeV protons and evaluate the likelihood that these sources could have produced the high Be/O observed in bodies accreted by polluted WDs.  

Using Saturn as a model, we find that rings composed mainly of water ice within the magnetosphere of a giant planet satisfy the constraints imposed by the excess Be concentrations exhibited by the polluted WDs. Mid-sized icy moons of Saturn evidently formed from rings \citep{Charnoz2009}, and indeed accretion of exomoons by WDs was anticipated. Moons stripped from their host planets were predicted to be a likely source of rocky/icy material for pollution of WDs based on an analysis of post-main-sequence scattering in WD planetary systems \citep{Payne2016a,Payne2016b}. It is worth remarking that the study of Li and Be as markers of stellar pollution by planets has a rich history \citep[e.g.,][]{Maia2019, Deliyannis1997}. We suggest that Be, and perhaps Li and B, excesses in WDs polluted by rocky and icy bodies are signatures of accretion of icy exomoons.

We focus our study on GALEX J2339-0424 as an example of a polluted WD with evidence for excess spallogenic nuclides, but our general conclusions also apply to GD 378. An evaluation of settling effects on estimates of the composition of the polluting parent body and a discussion of the duration of accretion for GALEX J2339-0424 is detailed in Section \ref{methods}. In Section \ref{discussion} we outline the proposed scenario for acquiring excess Be in view of the various alternatives. Section \ref{conclusions} provides a brief summary of our conclusions.

\section{The Parent Body Accreted by WD GALEX J2339-0424} \label{methods}
\subsection{The Effects of Settling on Element Ratios} \label{SettlingModel}
As reported by Klein et al. (2021), GALEX J2339-0424 exhibits significant pollution by the major and some minor and trace rock-forming elements (Table \ref{table}). The uncertainties reported in Table \ref{table} represent the  spread in values obtained from different transition lines for the same element (see Klein et al., 2021, for a more detailed analysis). At face value, the composition of the rocky and icy material comprising the pollutants for GALEX J2339-0424 include an excess of oxygen relative to the other rock-forming elements suggestive of a large volume fraction of water ice and an excess in Be relative to chondritic abundances by a factor of more than $\sim 500\times$ (in CI chondrites in the solar system, the Be/Fe atomic ratio is $7.3 \times 10^{-7}$; \citeauthor{Lodders2019} \citeyear{Lodders2019}). In detail, elemental concentrations for the accreted parent body material are extrapolated from the WD photospheric abundances by taking into account changes in element ratios produced by diffusion out of the stellar atmosphere together with the flux of material accreting onto the surface of the WD. In general, three different phases of accretion/diffusion are recognized for pollution of WDs: a build-up phase, a steady-state phase, and a declining phase \citep[e.g.,][]{Dupuis_1992,Dupuis_1993,RN50}. Differences in diffusive velocities will modify abundance ratios in the second two phases, imparting disparities between the relative elemental abundances in the accreted body and those in the atmosphere of the WD. Generally, heavier elements sink faster than lighter elements, but there are some exceptions. 

Here we use the model from \cite{Jura2009} for the time-dependent mass of element $z$ in the WD convective layer ($M_{{\rm CV},\it{z}}(t)$) assuming that the mass of the debris disk feeding the surface of the star decays exponentially as settling through the convective layer proceeds. This model simulates all three phases of accretion with time. The solution for the time-dependent mass of element $z$ in the convective layer is

\begin{align}
   M_{\rm CV,\it{z}}(t) = \frac{M^{\rm o}_{\rm PB,\it{z}}\tau_z}{\tau_{\rm disk}-\tau_z} \left [ e^{- t/\tau_{\rm disk}} - e^{- t/\tau_z} \right ],
   \label{Jura_soln}
\end{align}

\noindent where $M^{\rm o}_{\rm PB,\it{z}}$ is the initial mass of $z$ in the parent body that forms the circumstellar disk, $\tau_{\rm disk}$ is the e-folding time for the depleting disk mass of parent body material, and $\tau_z$ is the e-folding time for diffusive settling of element $z$ through the WD's convective zone. We use this model to explore the effects of elemental settling through the WD envelope on calculated element ratios as a function of time.

\begin{figure}
\begin{center}
 \includegraphics[width=3.5 in]{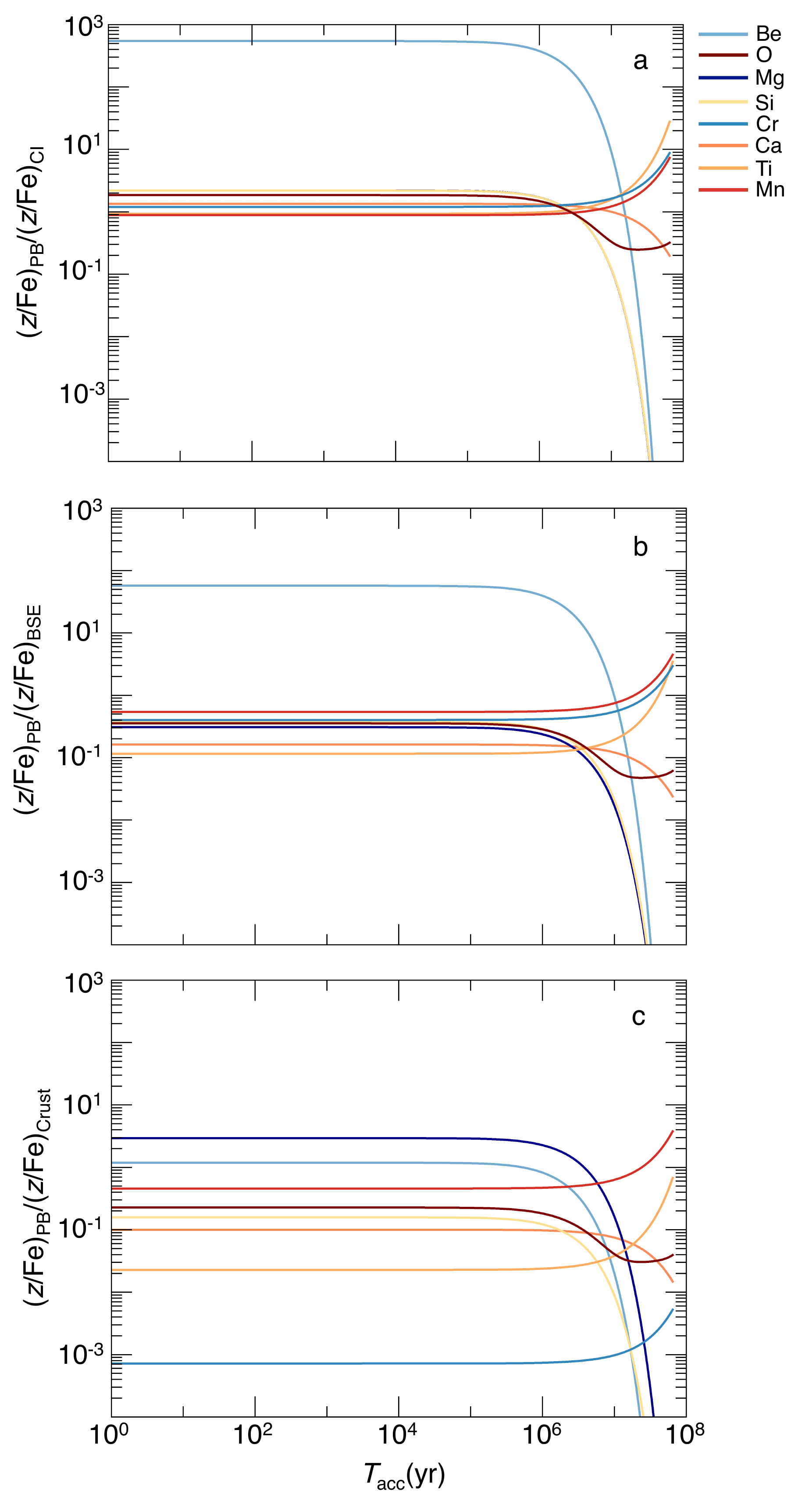}
 \caption{Element/iron atomic ratios, $z$/Fe, for the parent body accreted by GALEX 2339-0424, relative to $z$/Fe in CI chondrite, bulk silicate Earth and the Earth's average continental crust, as a function of the duration of the accretion event, $T_{\rm acc}$, calculated using Equation~\ref{Jura_solved} assuming $\tau_{\rm disk}$ = $10^5$ yr. We compare the calculated parent body elemental abundances accreted by GALEX 2339-0424 to CI chondrite \citep[a,][]{Lodders2019}, bulk silicate Earth (BSE) \citep[b,][]{RN28}, and the Earth's continental crust \citep[c,][]{Rudnick2014}, each having distinctive Be abundances relative to the other major elements. Note that the curves for Mg and Si overlap in panel a. The best-fit composition for the parent body accreting onto GALEX 2339-0424 is CI chondrite, with Be being anomalously high by two orders of magnitude.}
   \label{crossing}
\end{center}
\end{figure}

Differential settling through the WD envelope may cause lighter elements to appear in excess, altering the geochemical interpretation \citep[e.g.,][]{Doyle2020}. In particular, because Be is among the lightest metals discovered in a polluted WD, we require an evaluation of whether a high beryllium concentration in the atmosphere of the WD could be due simply to the higher rates of gravity-driven settling for heavier elements, compared to Be. In practice, we calculate element abundances relative to Fe in the WD atmosphere, as a function of time. We assume a CI chondrite composition for the parent body on a water-free basis, only including the oxygen that was available to pair with the other rock-forming elements, including Be, in the parent body. As an example, if a CI chondrite accreted onto a WD similar to GALEX J2339-0424, we would expect to see Be/Fe ratios elevated to 500 times chondritic by $\thicksim$ 15.1 Myr. At that time, Al/Fe, Mg/Fe and Si/Fe would be $\thicksim$ 65, $\thicksim$ 82, and $\thicksim$ 80 times chondritic, respectively. The abundance of Al is not constrained, but Mg and Si are observed not to be supra-chondritic in GALEX J2339-0424, showing that preferential settling of heavier elements cannot explain the high Be/Fe ratio in this WD, and that the debris disk material feeding the WD must itself have elevated Be abundances. Another possibility is that Be accumulated over time from multiple accretion events. However, it is straightforward to show that if excess Be was a residue of preferential settling of heavier elements left over from a parent body from an earlier accretion episode, one should expect excesses in Al, Mg and Si relative to Fe as well.

One can calculate the composition of the accreted parent body as a function of the duration of the accretion event, $T_{\rm acc}$, assuming a value for $\tau_{\rm disk}$ and an exponentially decaying debris disk. Solving Equation \ref{Jura_soln} for $M^{\rm o}_{\rm PB,\it{z}}$ where  $t = T_{\rm acc}$, yields

\begin{align}
    {M^{\rm o}_{\rm PB,\it{z}}(T_{\rm acc})} = \frac{M_{\rm CV,\it{z}}\left [\tau_{\rm disk}-\tau_z \right ]}{\tau_z \left [ e^{- T_{\rm acc}/\tau_{\rm disk}} - e^{- T_{\rm acc}/\tau_z} \right ]}.
   \label{Jura_solved}
\end{align}

\noindent We calculate the mass of each element in the convective zone, $M_{\rm CV, \it{z}}$, by using the mass of the convective layer, obtained from $\log(M_{\rm CV}/M_{\rm WD})$ in Klein et al. (2021), and converting number ratios, $z/{\rm He}$, to mass ratios. For the purposes of this work, in the first instance we assume that $\tau_{\rm disk} = 10^5$ yr. These inferred abundances can then be compared to hypothetical starting compositions. 

Figure \ref{crossing} shows an example calculation for the parent body compositions assuming different durations, $T_{\rm acc}$, for the accretion event. We compare inferred abundances of $\it{z}$/Fe in GALEX J2339-0424, to $\it{z}$/Fe in CI chondrite, bulk silicate Earth and continental crust \citep{Lodders2019,RN28,Rudnick2014}. In these calculations we again calculate the parent body accreted by GALEX J2339-0424 and the comparison rocks on a water-free basis, excluding the excess oxygen in the WD that would have existed as water ice in the parent body. Values of unity for the ordinate in Figure \ref{crossing} indicate a match between the calculated composition of the parent body and the reference rock material for the accretion duration indicated on the abscissa if uncertainties are well characterized. We conclude that the composition of the planetary materials are like CI chondrite, and that the Be abundance is simply anomalously high in GALEX J2339-0424. All the other elements' predicted abundances match those in the white dwarf's atmosphere to within a factor of 2 or less, especially if the accretion has been ongoing for about 2 to 3 Myr. This strongly suggests that the accreted body was a chondrite-like body similar to those in our solar system. This composition is consistent with the calculated oxygen fugacity for the accreted body. Based on the mole fraction of FeO we calculate an oxygen fugacity expressed as the difference in $\log_{10} f_{\rm O_2}$ from that of the iron-w\"{u}stite reference, $\Delta$IW, of $-1.35$, similar to that of carbonaceous chondrites in general. The high concentration of Be stands out as the anomaly in being  $\thicksim$ 2 orders of magnitude more abundant in GALEX J2339-0424 than in a CI chondrite.

The failure of other relevant geological materials to fit the observed relative concentrations of the rock-forming elements in the WD demonstrates that the excess of Be cannot be explained by geochemical processes that might concentrate Be. The bulk silicate Earth is a poor match for the data: all elements would be under-abundant relative to Fe by factors of up to 5; and Be again would be overabundant by about 2 orders of magnitude if the accreted body was similar to bulk silicate Earth in composition. Beryllium tends to concentrate in the continental crust on Earth.  The composition of the Earth's continental crust also does not match the composition of the atmosphere of GALEX J2339-0424: although a reasonable match to the Be/Fe ratio can be made, the other ratios fail to match to even the order of magnitude level. Although unlikely to be major contaminants, we also performed similar calculations using accretion of the Be-rich mineral beryl (${\rm Be}_{3}{\rm Al}_{2}{\rm Si}_{6}{\rm O}_{18}$, e.g., aquamarine or emeralds) and other Be-rich minerals or rocks (e.g., pegmatites). These also failed to match the composition of the white dwarf atmosphere nearly as well as CI chondrite material.    

This chondrite-like parent body was water ice-rich. Three-quarters of the oxygen comprising the parent body accreted by GALEX J2339-0424 was in excess of that required to form the oxides of the rock-forming elements. The excess oxygen was presumably accreted as water ice. Therefore, the parent body that accreted onto GALEX J2339-0424 was approximately 85\% water by volume.

\subsection{Duration of Accretion and Mass of the Parent Body}
\begin{figure}
\begin{center}
 \includegraphics[width=3.5 in]{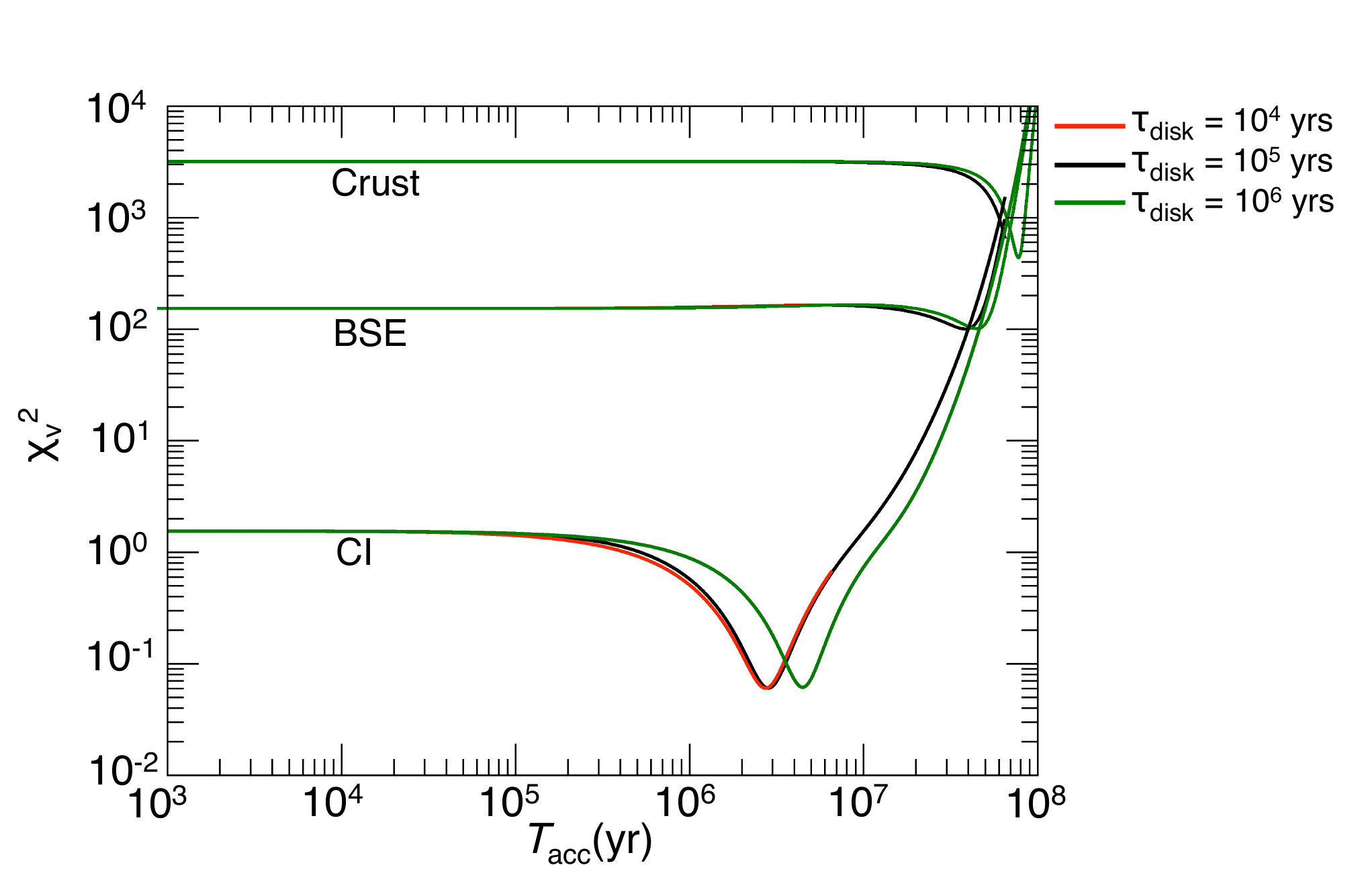}
 \caption{Reduced chi-squared, $\chi_{\nu}^2$, for fits of the parent body composition to average continental crust, bulk silicate Earth, and CI chondrite, as functions of the duration of the accretion to WD GALEX 2339-0424, $T_{\rm acc}$. Variations in concentrations in the atmosphere of the WD as functions of accretion duration are obtained using Equation \ref{Jura_solved}. Various disk e-folding timescales, $\tau_{\rm disk}$, are shown for comparison. The fits at each timescale for accretion are obtained for the major rock-forming elements Mg, Si, Fe, and Ca and the minor elements Ti and Mn. The best fit is obtained for a CI chondrite composition and timescales for accretion of between 2.4 and 4.0 Myr, as indicated by the minima in the reduced chi-squared value relative to CI chondrite for different values for the lifetime of the debris disk. The fits for both bulk silicate Earth and continental crust are sufficiently poor that these compositions can be excluded.}
   \label{Chi2}
\end{center}
\end{figure}

Under the hypothesis that the composition of the parent body is like that of a CI chondrite, except for the concentration of Be, we can estimate the timescale for accretion and settling onto GALEX J2339-0424. We do this by searching for the best fit between the parent body element ratios and CI chondrite element ratios, as a function of the duration of accretion and settling. We search for the value of $T_{\textrm{acc}}$ that yields a minimum in the reduced  chi-squared statistic, $\chi_{\nu}^2$, to assess the most likely timescale for accretion and settling for the parent body accreting onto GALEX J2339-0424. In our analysis we make use of the random errors, $\sigma_{\rm spread}$, listed in Table \ref{table}, and exclude the correlated systematic errors associated with the effective temperature and gravity of the host WD; a more detailed error analysis is outlined in Klein et al. (2021). In general, shorter timescales are better fits than much longer timescales, and we find the minimum in the reduced chi-squared statistic occurs for an accretion time of $\approx 2.5$ Myr, where $\tau_{\rm disk}$ = $10^5$ yr is assumed and the debris disk decays exponentially (Figure \ref{Chi2}). Adopting this timescale, $T_{\rm acc}$, the inferred mass for the parent body accreted by GALEX J2339-0424 is $4 \times 10^{23}$ g, or $\thicksim$ 1/2 the mass of Ceres, and Be is in excess relative to chondritic abundances by a factor of $\thicksim 200\times$. If we instead assume $\tau_{\rm disk}$ = $10^4$ or $10^6$ yr, the minimum value of $\chi_{\nu}^2$ occurs for accretion durations of $\approx 2.4$ and $4.0$ Myr, respectively. The best-fit $\chi_{\nu}^2$ value is $<$1, suggesting an over-estimation of uncertainties in the elemental ratios, but also indicating the goodness of the fit to a CI chondrite composition. For comparison, we also show $\chi_{\nu}^2$ for bulk silicate Earth and continental crust (Figure \ref{Chi2}). The high  $\chi_{\nu}^2$ values underscore that these compositions are not adequate matches to the parent body accreted by GALEX J2339-0424. The volume fraction of water obtained for the parent body from the best-fit is approximately 85\%, similar to the value obtained from the uncorrected data.

\begin{figure}
\begin{center}
 \includegraphics[width=3.4 in]{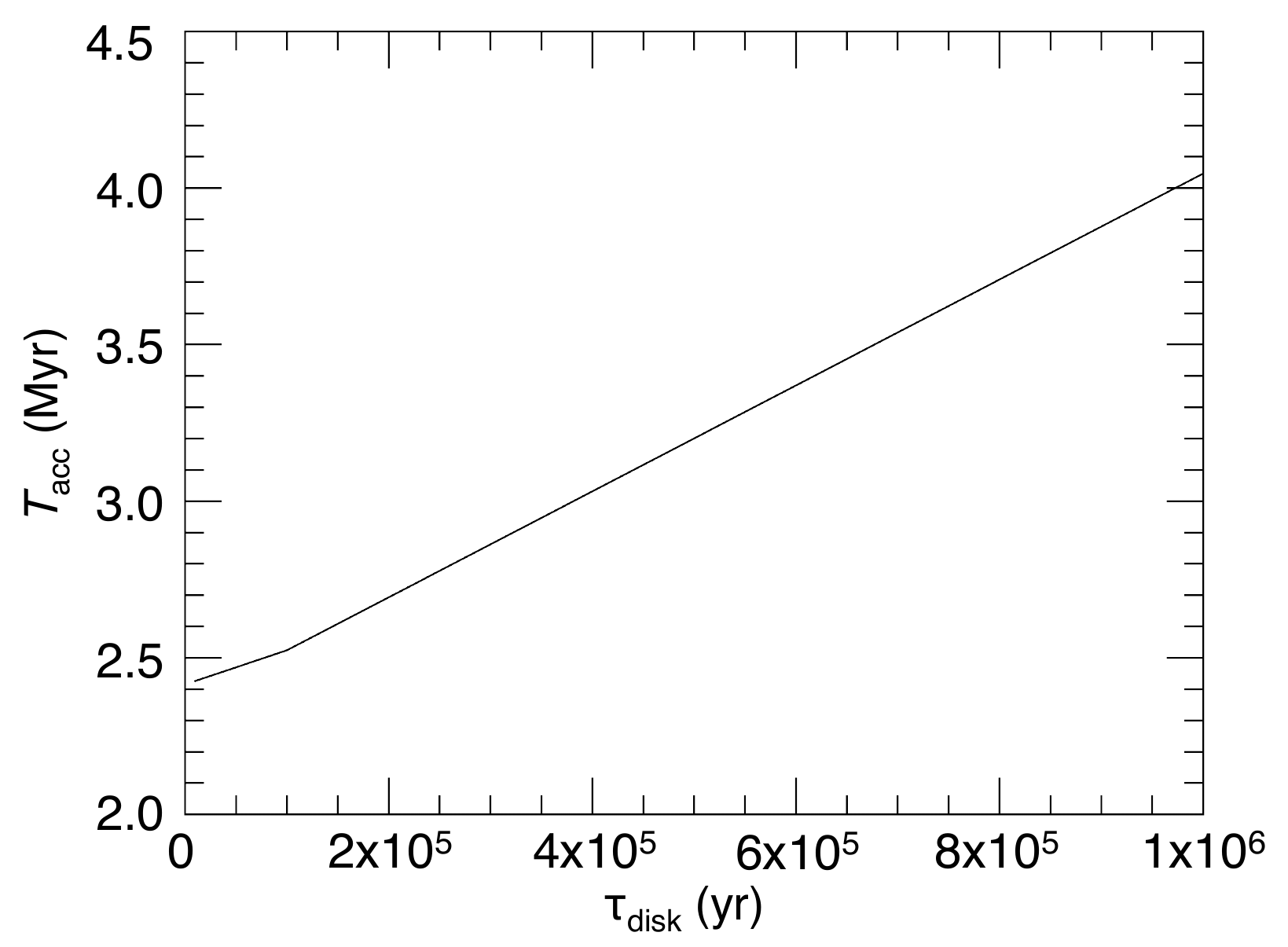}
 \caption{Relationship between the optimal accretion duration, $T_{\rm acc}$, as indicated by the minima in $\chi_{\nu}^2$ versus $T_{\rm acc}$ in Figure \ref{Chi2}, and the assumed debris disk e-folding timescale, $\tau_{\rm disk}$.}
   \label{Chi2extrap}
\end{center}
\end{figure}

The dependence of the derived accretion duration on the assumed value of $\tau_{\rm disk}$ is shown in Figure \ref{Chi2extrap}. Assuming that $10^4$ to $10^6$ years spans the likely values for the e-folding time for the debris disk, we conclude that the accretion event that added the rock-forming elements to GALEX J2339-0424 lasted for 2 to 4 Myr, and so the mass of the accreted parent body was $3\times 10^{23}$ to $1\times 10^{24}$g. Assumed durations for accretion less than 2.5 Myr would decrease the estimated mass of the parent body.

\section{Source of Beryllium excess} \label{discussion}
\subsection{Constraints on the Radiation Environment}

The excess Be observed in this polluted WD is almost certainly due to spallation of heavier nuclei (in particular, O) in rock or ice since it cannot be explained by differential settling in the atmosphere of the WD nor by geochemical processes. Additionally, winds from the WD itself would only be efficacious if the star were rapidly rotating, or another mechanism such as a magnetic field were available to capture protons. GALEX J2339-0424 is neither magnetic nor rapidly rotating. In order to determine the radiation environment in which the accreted parent body formed, we require an environment that can produce the observed Be/O number ratio of approximately 10$^{-5}$. This ratio is relatively insensitive to the details of the settling history (e.g., Figure \ref{crossing}). 

In order to estimate the proton fluence required to explain the observed Be/O ratio, we consider a first order rate equation for the spallation production of Be:

\begin{align}
    \frac{d\nBe}{dt}&=kn_{\rm p} \nO\\
    &=\sigma f_{\rm p} \nO \notag ,
\end{align}

\noindent where the product of the proton number density ($n_{\rm p}$) and rate constant ($k$) is replaced by the cross-section for the spallation reaction ($\sigma$) and the proton flux ($f_{\rm p}$). Assuming no initial Be at time zero, a reasonable approximation given the magnitude of the excess in Be observed, integration yields

\begin{align}
    \label{nBenO}
    \frac{\nBe}{\nO}&= \sigma f_{\rm p} \Delta t \\
    &= \sigma F_{\rm p}.   \notag
\end{align}

\noindent Here the proton fluence ($F_{\rm p}$) indicated by the Be/O number density ratio provides the constraint on the radiation environment. The cross section for Be production by the reaction $^{16}$O(p,X)$^{9}$Be is $\thicksim$ $10^{-26}$ cm$^{2}$ \citep{Moskalenko2003} with a minimum required energy of about 10 MeV. The proton fluence required to obtain the observed Be/O atomic ratio in the accreted material is therefore

\begin{equation}
    F_{\rm p} \thicksim \frac{10^{-5}}{10^{-26}\,{\rm cm}^2} \thicksim 10^{21}\,{\rm cm}^{-2}.
    \label{eq:fluence}
\end{equation}

\noindent The cross sections for Li and B production are comparable to the cross section for production of Be, and would yield similar Li/O and B/O ratios.  

Endeavors to explain the origin of the short-lived radioisotope $^{10}$Be in calcium-aluminum-rich inclusions (CAIs) formed in the early solar system \citep[e.g.,][]{Mckeegan2000} have given rise to a significant literature on $^{10}$Be production by spallation. The findings of these studies provide useful constraints on various astrophysical environments for the formation of not only $^{10}$Be, but also as a corollary, for the formation of the isotopes of Li, Be, and B in general. These findings can be summarized as referring to three distinctive environments and/or processes for the formation of spallogenic light nuclides. These include production of ${}^{10}{\rm Be}$ atoms in star-forming molecular clouds by spallation by GCRs accelerated by core-collapse supernovae (CCSNe) \citep{Desch2004,Tatischeff2014}, enrichment from a single low-mass CCSN adjacent a region of star formation \citep{Banerjee2016}, or irradiation of the inner edge of the protoplanetary disk by stellar energetic particles (SEPs) from the young star \citep{RN532,Gounelle2006,Jacquet2019}. In the case of CAIs in the early solar system, precise isotopic ratios, including $^{10}$Be/$^{9}$Be, are brought to bear in order to evaluate the efficacy and plausibility of these various suggested environments for the spallation reactions. In the case of a polluted WD, we do not have access to isotope-specific data. Therefore, we make use of the fluence indicated by Equation \ref{eq:fluence} as the primary arbiter for the environment that formed the observed excess in Be (and by inference, Li and B as well).

The flux of ambient Galactic cosmic rays (GCRs) with sufficient energy ($\thicksim$ 10 MeV/nucleon) to induce spallation reactions to form Be in the solar neighborhood is $\thicksim$ 1 to 10 protons cm$^{-2}$ s$^{-1}$ \citep[e.g.,][]{Tatischeff2014}. To reach a fluence of 10$^{21}$cm$^{-2}$ that flux would have to act for 10$^{12}$ to 10$^{13}$ years, an impossibly long timescale. A larger flux of protons is required. Core-collapse supernovae are one exogenous source of high proton flux. The energy fluence ($F_{\rm E}$) required for Be production relative to oxygen is obtained from the product of the 10 MeV minimum energy per particle and the proton fluence, yielding $10^{16}$ erg cm$^{-2}$. Based on the typical non-neutrino energy of a SN of $10^{51}$ erg, we can write the energy fluence due to all particles, and light, as

\begin{equation}
    F_{\rm E}=3\times 10^{12} \left(\frac{\eta}{0.1}\right) \left(\frac{E_{\rm SN}}{10^{51}\,{\rm erg}}\right)
    \left( \frac{\rm r}{1 \,{\rm pc}} \right)^{-2} {\rm erg \,cm^{-2}},
\end{equation}

\noindent where $\eta$ is the fraction of the SN energy carried by protons that produce Be, which we have arbitrarily scaled to 0.1. Therefore, if 10\% of the total energy of the SN remnant went towards the production of Be (similar to the fraction of kinetic energy converted to escaping accelerated particles, \citealt{Tatischeff2014}), the energy fluence necessary to produce the observed Be/O ratio would require the SN source to be 0.025 pc from the planetary system. Besides being exceptionally improbable, at these distances, the system is unlikely to survive the CCSN event  \citep{Portegies2018}.  

A similar argument applies for the potential production of spallogenic nuclides as a result of winds from Wolf-Rayet stars (WR) in massive star-forming regions like Orion \citep[e.g.][]{Ramaty1998,Majmudar1999,Kozlovsky1997}. In this case, energetic $^{12}$C and $^{16}$O comprising the WR winds experience spallation upon striking protons in the ISM or the protoplanetary disk \citep{Kozlovsky1997,Prantzos2012}. We can assess the likelihood that this reverse process is important for the formation of Be by examining the average energy per O and C emitted. This energy is obtained using $E_{\rm winds}(m_{\rm avg}/M_{\rm C + O})$ where $E_{\rm winds}$ is the total energy released by the WR winds integrated over the lifetime of the WR phase, $m_{\rm avg}$ is the weighted mean mass of $^{12}$C and $^{16}$O nuclides (g/atom), and $M_{\rm C + O}$ is the mass of $^{12}$C and $^{16}$O released (g). For a typical WR lifespan (through the WC or WO phase) of $\sim 5\times 10^5$ yr, and a maximum wind power of $10^{38}$ erg s$^{-1}$ \citep[e.g.][]{Prajapati2019}, one obtains $E_{\rm winds} \sim 1\times 10^{51}$ erg, comparable to $E_{\rm SN}$. Mass loss rates for WR stars are $10^{-5}$ $M_\odot$ yr$^{-1}$ \citep{Crowther2007} and with the total fraction of $^{12}$C + $^{16}$O being on the order of $0.65$ \citep[e.g.][]{Sander2020,Tramper2013}, the mass of C and O released is about $3.5$ $M_\odot$. Since C/O is $\gg 1$ in the winds, $m_{\rm avg} \sim 12$. Using these values we find that the energy per C and O nuclei for the WR winds is $\sim 2$ MeV. If we reduce the timescale for the WR phase to $1\times 10^{5}$ yr, we obtain $\sim 10$ MeV per C and O nuclide. As no additional efficiency or dilution factors have been included, this result is something of a maximum, and we take this as indication that WR winds are only marginally capable, at best, of producing the fluence of $>$ 10 MeV C and O nuclei required to generate significant excesses in Li, Be and B by spallation reactions.

Accumulations of Li, Be, and B produced by low-mass CCSNe by neutrino-driven reactions like $^{12}$C($\nu,\nu^\prime  {\rm pp}$)$^{10}$Be are also feasible. However, the $^9$Be/$^{16}$O production ratio for the low-mass (12 M$_\odot$) CCSN progenitor advocated by \cite{Banerjee2016} is $4\times 10^{-10}$. Because the Be/O ratio is lower for larger CCSNe \citep{Banerjee2016}, this Be/O low-mass CCSN production ratio represents a maximum. This is already orders of magnitude lower than the Be/O $\sim 1 \times 10^{-5}$ observed in GALEX J2339-0424. Furthermore, this injected supernova material would be diluted with oxygen in the planetary system. The abundances of short-lived radionuclides like ${}^{26}{\rm Al}$ in the solar nebula suggest dilution factors of 4 to 5 orders of magnitude. The mass of Be produced by this mechanism is far too small in comparison to oxygen to account for the observation in GALEX J2339-0424. 

\begin{figure*}
\begin{center}
 \includegraphics[width=\textwidth]{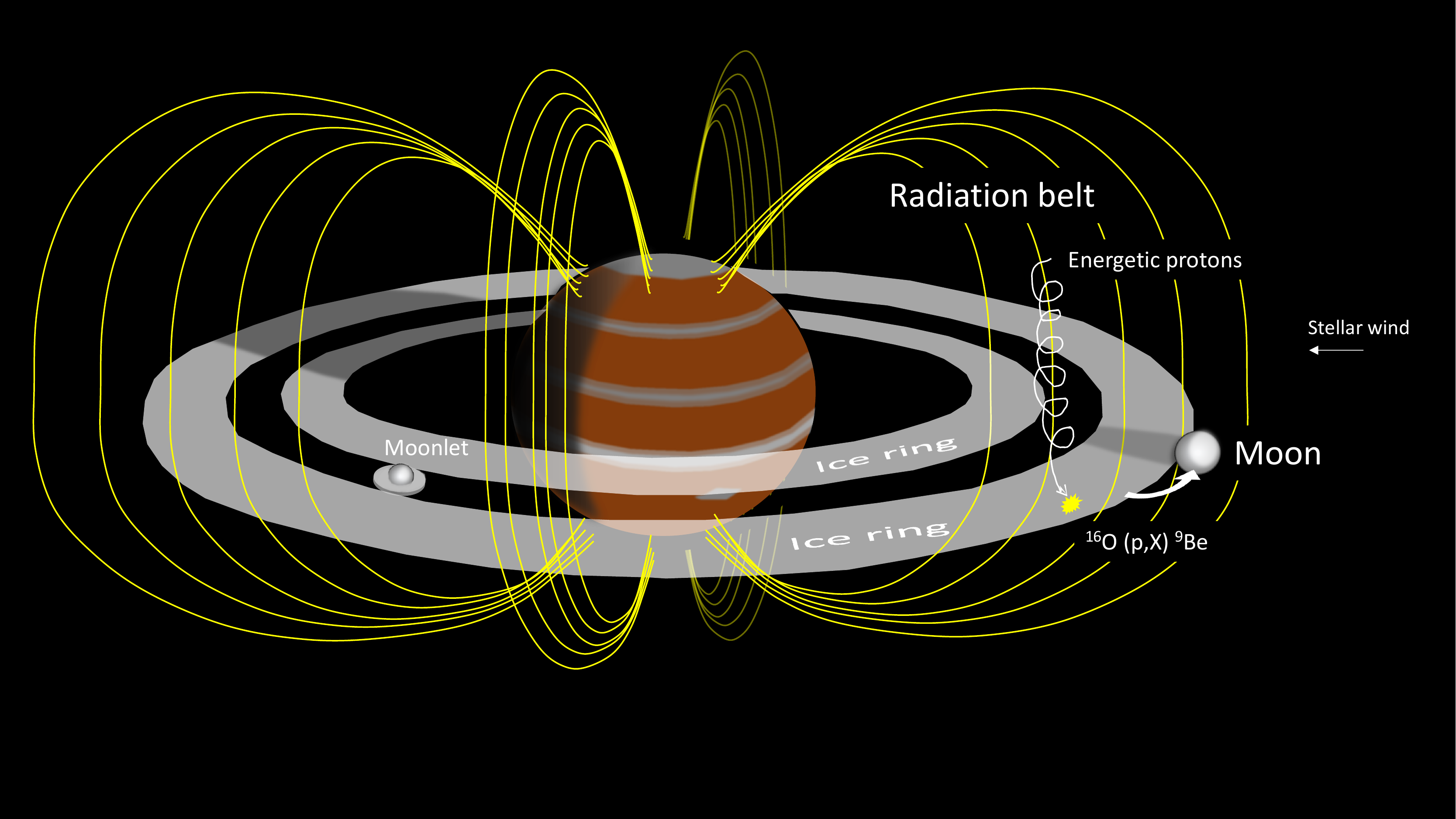}
 \caption{Schematic diagram depicting the proposed environment for formation of ices enriched in spallogenic nuclides. The source of the trapped magnetospheric particles is external, mainly from stellar winds. Once trapped, high-energy protons mirror along magnetic field lines until they interact with icy material in the ring by the reaction $^{16}$O(p,X)$^9$Be. The mass of the rings is transferred back and forth from fine particles to moonlets, until eventually icy ring material accretes around a rocky core and a moon is formed at the outer edge of the disk that includes the product $^9$Be \citep{Cuzzi2009,Charnoz2009,Charnoz2011}.}
   \label{diagram}
\end{center}
\end{figure*}

In contrast to these exogenous sources, the fluence of energetic protons emanating from a protostar in its first $\thicksim 10$ Myr, during the lifetime of its surrounding protoplanetary disk, far exceeds those of normal GCRs integrated over the 10 Gyr age of the Galaxy; energetic proton fluxes from young stars at 1 au are about 10$^7$ times the GCR fluxes. From \cite{RN532} we can estimate the fluence of SEPs with energies $>$ 10 MeV/nucleon using

\begin{equation}
    {F_{\rm p}} = \frac{L_{\rm p}}{L_{\rm X}}\frac{L_{\rm X}}{L_{\rm star}}\frac{L_{\rm star}}{\rm 4 \pi {\it r}^2}\Delta t, 
    \label{youngstareqn}
\end{equation}

\noindent where $L_{\rm p}/L_{\rm X}$ is the proton luminosity ($L_{\rm p}$) scaled to X-ray values ($L_{\rm X}$), which at the peak of a G-star spectrum ($\thicksim$ 10 MeV) is 0.09 \citep{Lee1998}. For G stars, $L_{\rm X}$ $\thicksim 6\times 10^{30}$ erg s$^{-1}$ in the first 10 Myr of the stellar lifetime \citep{Feigelson1982}. Therefore, a young solar-mass star would have a proton luminosity at 10 MeV of $L_{\rm p} = 5\times 10^{29}$ erg s$^{-1}$. The progenitor main sequence star for GALEX J2339-0424 was likely $\thicksim$ 1.5 ${\rm M}_{\odot}$ \citep{Cummings2018}, such that $L_{\rm star} \thicksim {5\,L}_{\odot}$ ($1\,L\odot = 3.8\times 10^{33}$ erg s$^{-1}$). Using 10 MeV as the kinetic energy of the protons ($1.6\times 10^{-5}$ erg), scaling for a 1.5 ${\rm M}_{\odot}$ star, and adjusting the flux of protons for a spherical geometry at a distance from the central star of $r \approx$ 1 au, the flux of protons, $f_{\rm p}$, is $\thicksim 6\times 10^7$ protons cm$^{-2}$ s$^{-1}$. Assuming irradiation of the protoplanetary disk lasts approximately 5 Myr, the fluence, $F_{\rm p}$, is then $\thicksim 9\times 10^{21}$ protons cm$^{-2}$. Therefore, stellar winds early in the history of the planetary system are  in principle a feasible source of high-energy protons with the fluence required by the Be/O ratio observed in this polluted WD. 

However, the energy loss of protons due to ionization of hydrogen severely limits Be production in the presence of a protoplanetary gas. The stopping density for 10 MeV protons in hydrogen gas is on the order of 70 g cm$^{-2}$ \citep{Clayton1995}. For typical inner-disk midplane mass densities of 10$^{-10}$ g cm$^{-3}$ the stopping distance for the relevant incident protons is approximately 70 g cm$^{-2}/10^{-10}$ g cm$^{-3}$ = 7$\times 10^{11}$ cm, or 0.05 au. This limits the region of sufficient irradiation to within 0.05 au into the inner edge of the disk. Such a localized environment for rock formation makes this scenario unlikely, especially in view of multiple instances of accretion of Be-rich rocky bodies.

\subsection{Spallation in the Radiation Belts of Giant Planets and Brown Dwarfs}

Based on the discussion above, the observed excess Be found in GALEX J2339-0424 appears to require that the accreted parent body formed in a local region of unusually high proton flux that was largely free of hydrogen gas. Radiation belts around giant planets satisfy these conditions. 

Charged particles (mostly protons and electrons) from the solar wind can become trapped and forced to gyrate around the magnetic field lines of a giant planet, eventually mirroring back and forth between the magnetic poles and filling the planet's magnetosphere with energetic particles \citep[e.g.,][]{VanAllen1980}. On Earth, the magnetosphere traps solar wind particles, preventing them from reaching the atmosphere, except when a contraction of the magnetic field lines causes the particles to precipitate in the atmosphere and form aurorae. Similarly, Jupiter and Saturn have radiation belts of trapped energetic particles mirroring from pole to pole that have been recorded by spacecraft \citep[e.g.,][]{Bolton2007,Cooper2018}, as well as aurorae \citep[e.g.,][]{Nichols2014,Nichols2017}. In general, the rings of the giant planets, including Saturn, lie within the magnetospheres of the host planet and are subject to irradiation by energetic particles in these radiation belts. The irradiation of ice particles in a giant planet's rings is depicted in Figure~\ref{diagram}.

In order to assess the plausibility of this environment for explaining the Be excess observed in GALEX J2339-0424, we evaluate irradiation timescales required by the data using Equation \ref{nBenO}, but modified to include the fraction of oxygen present as ice in the rings that will be subject to irradiation. The stopping power of water ice is $\approx 40$ to $10 \, {\rm MeV} / ({\rm g} \, {\rm cm}^{-2})$ for energetic protons of energies 10 MeV and 100 MeV, respectively \citep{Berger2017}. Protons of these energies would be stopped completely by column densities $\Sigma_{\rm stop}$ of $0.25 \, {\rm g} \, {\rm cm}^{-2}$ to $10 \, {\rm g} \, {\rm cm}^{-2}$, respectively.  The corresponding stopping lengths,  based on the density of water ice, are 0.27 cm and 11 cm, respectively. 

The mass of Saturn's rings of $\approx 1.5 \times 10^{22}$ g \citep{Iess2019} and the area of the rings of $\approx 5 \times 10^{20}$ ${\rm cm}^2$ \citep{Charnoz2009} suggest that the water ice column density in the rings, $\Sigma_{\rm rings}$, is of order $\sim 30$ g ${\rm cm}^{-2}$. This is about 3 times the maximum stopping distance for the energetic protons, suggesting that protons are efficiently stopped by the ring ices. Before these particles are stopped, they have the opportunity to spall O and create Be nuclei. The ratio $\Sigma_{\rm stop}/\Sigma_{\rm rings}$ is a dilution factor for the production of Be relative to oxygen where $\Sigma_{\rm stop}/\Sigma_{\rm rings} \le 1$. We modify Equation \ref{nBenO} to include this dilution factor: 

\begin{equation}
    \frac{\nBe}{\nO}= \sigma f_{\rm p} \Delta t \frac{\Sigma_{\rm stop}}{\Sigma_{\rm rings}}, \  \  \  \  \Sigma_{\rm stop}/\Sigma_{\rm rings} \le 1.
    \label{eq:nBenO_modified}
\end{equation}

\noindent We note that the fraction of energetic protons by reactions that form Be nuclei is $(\Sigma_{\rm stop} / 18 m_{\rm p}) \sigma$ where $m_{\rm p}$ is the mass of a nucleon.  This fractional factor $\xi$ is $\approx 8 \times 10^{-5}$ to $3 \times 10^{-3}$ for protons with energies of 10 MeV and 100 MeV, respectively. We assume this fraction $\xi$ of protons that spall ices to form Be in the rings is a robust property of the system; even the smaller particles in Saturn's rings are cm to meters in size  \citep{Cuzzi2009}, comparable to, or a few times larger than, the 0.25 to 10 cm stopping distances of 10 MeV to 100 MeV protons in water ice. 

The present-day flux of MeV protons in Saturn's magnetosphere is measured to be $6 \times 10^4 \, {\rm cm}^{-2} \, {\rm s}^{-1}$ \citep{Kollman2015}. For $\Sigma_{\rm stop}/\Sigma_{\rm rings} = 1/3$, Equation \ref{eq:nBenO_modified} shows that the energetic proton flux in the Saturnian radiation field, $f_{\rm p}$, corresponds to an implausibly long timescale of  $2 \times 10^9$ yr in order to produce the observed atomic Be/O ratio of $10^{-5}$ in the parent body accreted by GALEX J2339-0424. The $~$MeV proton flux  in the radiation belt of Jupiter is higher, with a value of about $10^7 \, {\rm cm}^{-2} \, {\rm s}^{-1}$ \citep{Sawyer1976}. This flux corresponds to a radiation timescale of $1 \times 10^7$  years.  Estimates for the residence time of ices in Saturn's rings are on the order of $10^7$ to $10^8$ years \citep{Charnoz2009}, suggesting that a Jovian-like radiation flux is a plausible source for the irradiation of ices comprising the parent body accreted by GALEX J2339-0424.  

The energetic proton flux in Equation \ref{eq:nBenO_modified} depends on the stellar wind intensity of the host star at the location of the planet, the efficiency with which the planet traps the particles, and the sink terms for protons. Trapping efficiency depends foremost on the magnetic field, which in turn depends on the mass of the planet, its rotation rate, and the conductivity of its interior. The rings are a significant sink for the protons, but not the principal determining factor for $f_{\rm p}$. The mass of Saturn's rings is $\sim 10^6$ times that of Jupiter's rings while the Saturnian radiation belt MeV proton flux is about $10^{-2}$  that of Jupiter.  The latter scales more closely with the $\sim$ 20-fold difference in magnetic moment for the two planets \citep{Went2011} than with their respective ring masses, indicating that the higher MeV proton flux in Jupiter's radiation belt compared with Saturn is primarily attributable to the higher Jovian magnetic field. This, in turn, suggests that Jovian-like proton fluxes are not precluded by the mere presence of rings for the planet that hosted the exomoon accreted by the white dwarf in this case.

Ices in a ring system of a giant planet around a relatively young ($< 10^8$ yr) protostar could easily develop the Be/O ratio of  $10^{-5}$ in $\sim 10$ Myr if they were irradiated for $\sim 10^7$ yr within the planet's magnetosphere. According to parameterizations by \cite{Sterenborg2011}, the X-ray luminosity of the Sun scales as $t^{-1.74}$, and the mass flux in the solar wind scales as $t^{-2.33}$, where $t$ is the time since the Sun formed. In its first tens of Myr, the solar wind easily could have been $\sim 10^6$ times stronger than today, and the X-ray luminosity and flux of energetic protons could have been $\sim 10^5$ times greater than today. A youthful system enhances the likelihood for the proton  fluence indicated by the observed excess in Be, although it is not required. 

Given that ices in the ring system within a giant planet's magnetosphere can develop a high Be/O ratio, we next address whether these ices could coalesce into a moon comparable in mass to the roughly $4 \times 10^{23} \, {\rm g}$ parent body accreted by GALEX J2339-0424. Such a body would be greater in mass than the Saturnian icy moons Mimas and Enceladus by factors of 10 and 4, respectively, but lower in mass than Tethys, Dione, and Rhea, by factors of 1.5, 3, and 6, respectively; the parent body accreted by the WD is comparable in mass to the icy Saturnian satellites. Moreover,  the density of the accreted body was $\approx 1.3 \, {\rm g} \, {\rm cm}^{-3}$ based on the fractions of CI-like rock and water ice indicated by the oxygen budget.  This density is comparable to the average densities of these icy moons (Mimas, $1.15 \, {\rm g} \, {\rm cm}^{-3}$; Enceladus, $1.61 \, {\rm g} \, {\rm cm}^{-3}$; Tethys, $0.98 \, {\rm g} \, {\rm cm}^{-3}$; Dione, $1.48 \, {\rm g} \, {\rm cm}^{-3}$; Rhea, $1.24 \, {\rm g} \, {\rm cm}^{-3}$). 

The origins of the Saturnian satellites are unclear, and many may be  primordial, but the innermost satellites are commonly hypothesized to have formed from the rings themselves. Previous models have suggested that Saturn's innermost, icy, moons formed as the rings viscously spread beyond the Roche limit, allowing the otherwise small (cm- to m-sized: \citet{Cuzzi2009}) particles to coalesce into a medium-sized moon \citep{Charnoz2009,Canup2010,Charnoz2010,Charnoz2011}. Indeed, modeling of the coupled tidal effects on orbital parameters and geophysical properties, by \cite{Neveu2019}, demonstrates that Mimas is almost certainly formed in the last 0.1 - 1 Gyr, presumably from the rings. These authors constrain the ages of the other moons to be much older, but given their common ice-rich compositions, it seems plausible that the other inner moons also formed from the rings, but much earlier. The model of \citet{Charnoz2011} predicts that moons form steadily over timescales from $\sim 10^6$ yr to $\sim 10^9$ yr, and formation timescales of $\sim 10^7$ years is reasonable.

We conclude that a mid-sized Saturnian-like icy moon has the right mass and composition to match the parent body accreted by GALEX J2339-0424, and that this body very plausibly could have formed with a high Be/O ratio due to the irradiation of the ices that comprised the moon's progenitor ring material by magnetospheric MeV protons (Figure~\ref{diagram}). This result is consistent with the prediction that icy exomoons liberated from their host planets are a likely source of WD pollution \citep{Payne2016a}.

Saturn provides a useful analog for the environment in which the body accreted by GALEX J2339-0424 formed. There are other analogs, however. As described by \cite{Kenworthy2015}, the 16 Myr-old, $0.9 \, M_{\odot}$ star 1SWASP J1407.93-394542.6 (``J1407") is orbited by a brown dwarf (BD) companion with an immense ring system, with a mass of $\sim 1 M_{\oplus}$ and extending out to a radius of 0.6 au. The surface density of the rings is therefore $\approx 25 \, {\rm g} \, {\rm cm}^{-2}$, remarkably similar to that of Saturn's rings. 

The presence of a large gap in the rings strongly suggests that a moon has already formed within these rings. The most probable mass of the BD companion, J1407b, is 13 to 26 Jupiter masses. It is unknown whether J1407b has an extensive magnetic field, but many BDs are magnetically active, with radio flares and aurorae \citep{Berger2001,Hallinan2007}. If J1407b has a magnetosphere like Jupiter's, it would extend out to $\sim 7 \times 10^6$ km, irradiating icy particles within it. Alternatively, the ices outside the magnetosphere would be directly irradiated by energetic particles emitted by the central star. Given the age of J1407, the flux of energetic protons would be about $2 \times 10^4$ times that from the Sun. Since J1407b orbits at about 3.9 au from its host star, the proton flux would be $\sim 10^6 \, {\rm cm}^{-2} \, {\rm s}^{-1}$, and Be/O ratios $\sim 1.4 \times 10^{-5}$ would be possible after irradiation for $\sim 10^8$ yr. We predict that rock or ices in the rings of J1407b have already acquired considerable amounts of spallogenic Li, Be, and B. 

\section{Conclusions} \label{conclusions}
GALEX J2339-0424 is a WD polluted by accretion of a parent body with inferred abundances of most elements conforming closely to a CI chondrite-like composition, but with a remarkable excess of Be (2 orders of magnitude more abundant than in a CI chondrite) and a considerable complement of water ice. We consider and rule out chemical fractionation processes as the cause of this enhancement. Based on an analysis of this WD as an archetypal example, we find that excesses in the spallation products Li, Be, and B in the parent body of rocky/icy debris accreted by a polluted white dwarf are most likely a signature of accretion of an icy exomoon formed around a giant planet. Other potential sites of spallation lack the fluence required to produce the observed excesses. 

The degree of enhancement of spallation products in an icy exomoon will depend on the flux of stellar energetic particles and the trapping efficiency of the planet's magnetosphere. The mid-sized moons of Saturn are close analogs to the inferred properties of the parent body polluting GALEX J2339-0424. The masses and densities of the Saturnian mid-sized icy moons are comparable to those determined for the icy parent body accreted by the WD.  A corollary of this study is that we predict that at least some of the mid-sized icy moons of Saturn (e.g., Mimas) should be enriched in Li, Be, and B. Additionally, the rings around the brown dwarf J1407b also would experience intense irradiation, and this system may also serve as an analog for how exomoons could form with elevated Be/O ratios. 

Ejection of icy exomoons (like Saturn's mid-sized moons) from giant exoplanets is considered a likely means of polluting WDs \citep{Payne2016a} after the central star evolves to a white dwarf. In the absence of viable alternative explanations, supra-chondritic Be/O ratios in polluted WDs may be a signpost of this process.

We focus on Be in this work, but we also expect to see WDs with overabundances of Li and/or B produced by the same processes. The detection of each of these elements depends on factors related to the effective temperature of the WD, $T_{\rm eff}$, and the resolution and wavelength range of the observations. Indeed, the recent report of Li in apparent modest excess of chondritic abundances relative to Ca in two ultra-cool WDs may be such a detection, although the authors in that study offered an alternative explanation for the high Li/Ca based on the age of these ancient stellar remnants \citep{Kaiser2020}.

\section*{Acknowledgements}
We thank Margaret Kivelson and Krishan Khurana, both of UCLA, for informative discussions about giant planet magnetospheres that helped shape the paper. We also thank Ben Zuckerman, of UCLA, for prompting our evaluation of Wolf Rayet stars. We would also like to thank the anonymous referee for their comments which improved the manuscript. This work was supported by NASA 2XRP grant no. 80NSSC20K0270 to EDY.


\bibliographystyle{aasjournal}
\bibliography{Beryllium.bib}

\end{document}